\documentclass[journal]{IEEEtran}

\ifCLASSINFOpdf

\else

\fi

\pdfoutput=1 

\usepackage{amssymb}
\usepackage{bm}
\usepackage{cancel}
\usepackage{algorithm}
\usepackage{algorithmicx}
\usepackage{algpseudocode}
\usepackage{amsfonts}

\usepackage[font=small,labelfont=bf,tableposition=top]{caption}
\usepackage[font=footnotesize]{subfig}
\usepackage{multicol}
\usepackage{blindtext}
\usepackage{float}
\usepackage{url}
\usepackage{cite}
\usepackage{color}
\usepackage[table]{xcolor}
\usepackage[pdftex]{graphicx}
\usepackage{epstopdf}
\graphicspath{./figures/}
\usepackage{pbox}
\usepackage{multirow}

\usepackage{titlepic}

\usepackage{algpseudocode}

\usepackage[cmex10]{amsmath}
\usepackage{array}

\newtheorem{mydef}{Theorem}

\usepackage{fixltx2e}

\begin{document}

\title{Empirically Estimable Classification Bounds Based on a Nonparametric Divergence Measure}

\author{Visar~Berisha, Alan~Wisler,
            Alfred O. Hero, and Andreas Spanias
\thanks{This research was supported in part by the Office of Naval Research grant N000141410722 (Berisha), National Institute of Health, National Institute on Deafness and Other Communicative Disorders Grant 1R21DC012558 (Berisha),Army Research Office grant  W911NF-11-1-0391 (Hero), National Science Foundation grant CCF-1217880, and National Institutes of Health (Hero) grant 2P01CA087634-06A2.}}



\def\Reals{\mbox{\rm I\kern-.2em R}} 

\maketitle

\begin{abstract}
Information divergence functions play a critical role in statistics and information theory. In this paper we show that a non-parametric $f$-divergence measure can be used to provide improved bounds on the minimum binary classification probability of error for the case when the training and test data are drawn from the same distribution {\em and} for the case where there exists some mismatch between training and test distributions. We confirm the theoretical results by designing feature selection algorithms using the criteria from these bounds and by evaluating the algorithms on a series of pathological speech classification tasks.
\end{abstract}


\IEEEpeerreviewmaketitle

\section{Introduction}

A number of information-theoretic divergence measures between probability distance functions have been introduced and analyzed in the literature \cite{bhattacharyya1946measure, kullback1951information, chernoff1952measure, hero01, Cha07}. They have been extensively used in many signal processing applications involving classification \cite{guorong1996}, segmentation \cite{hamza2003image}, source separation \cite{hild2001blind}, clustering \cite{banerjee2005clustering}, and other domains.

Among the different divergence functions, the family of $f$-divergences or Ali-Silvey distances  is perhaps the most widely used in signal processing \cite{csiszar04}. This family includes the total variation distance, the Bhattacharya distance \cite{bhattacharyya1946measure}, the Kullback-Leibler divergence \cite{kullback1951information}, and more generally, the Chernoff $\alpha$-divergence \cite{chernoff1952measure, hero01}. Because there exists an indirect relationship between the class of $f$-divergences and the minimum achievable error in classification problems \cite{ali1966general}, this family of divergence measures is particularly useful for this setting. Consider the problem of classifying a multi-dimensional feature vector, $\mathbf{x}$, into one of two classes, $\{0,1\}$. The conditional distributions are given by $f_0(\mathbf{x})$ and $f_1(\mathbf{x})$ and the prior probabilities are given by $p$ and $q$, respectively. The classifier that assigns a vector $\mathbf{x}$ to the class with the highest posterior is called Bayes classifier, and the error rate of this classifier is given by:
\begin{equation}
\epsilon^{\mathrm{Bayes}} = \! \! \! \! \! \! \! \! \! \! \! \! \! \int\limits_{p f_0(\mathbf{x}) \leq q f_1(\mathbf{x})} \! \! \! \! \! \! \! \! \! \! \! \! pf_0(\mathbf{x}) d\mathbf{x} \ \ + \! \! \! \! \! \! \! \!   \int\limits_{q f_1(\mathbf{x}) \leq p f_0(\mathbf{x})} \! \! \! \! \! \! \! \! \! \! \! \! qf_1(\mathbf{x}) d\mathbf{x}
\label{eq:BER}
\end{equation}
This is the minimum classification error rate, or the Bayes error rate (BER), that can be achieved by any classifier. Computing the BER requires evaluating the multi-dimensional integral over regions that can only be determined if one has perfect knowledge of the data distribution. As an alternative to computing the integral, a number of attempts have been made to bound this error using estimable measures of distance between probability functions \cite{chernoff1952measure, kailath1967divergence, hashlamoun1994tight, avi1996arbitrarily}.  

In this paper, we derive a new bound on classification error that is based on a nonparametric probability distance measure that belongs to the family of $f$-divergences. In the context of binary classification, this new measure has a number of appealing properties: (1) there exists an asymptotically consistent estimator of the divergence measure that does not require density estimates of the two distributions; (2) we show that there exists a local relationship between this new divergence measure and the Chernoff $\alpha$-divergence; (3) we derive tighter bounds on the BER than those based on the  Bhattacharya distance and derive empirical estimates of these bounds using data from the two distributions; (4) we derive bounds on the minimum achievable error rate for the case where training and test data in the classification problem come from different distributions. 

\subsection{Related work}

There are three lines of research that are related to the work presented in this paper:  information theoretic bounds on the Bayes error rate (and related quantities); bounds from the machine learning literature for the scenario where training and test data come from different distributions; and recent work on empirical estimates of the KL divergence.

The total variation (TV) distance is closely related to the Bayes error rate \cite{kailath1967divergence}. A number of bounds exist in the literature relating the KL divergence and the TV distance. The well-known Pinsker inequality provides a bound on the total variation distance in terms of the KL divergence \cite{csisz1967information}. Sharpened inequalities that bound the KL divergence in terms of a polynomial function of the TV distance were derived in \cite{kullback1967lower}. One drawback of the Pinsker-type inequalities is that they become uninformative for completely separable distributions where the KL divergence goes to $\infty$ (since the TV distance is upper bounded). Vajda's refinement to these bounds addresses this issue \cite{vajda1970note}. 

For classification problems, the well-known upper bound on the probability of error based on the Chernoff $\alpha$-divergence has been used in a number of statistical learning applications \cite{chernoff1952measure}. The tightest bound is determined by finding the value of $\alpha$ that minimizes the upper bound. The Bhattacharya (BC) divergence, a special case of the Chernoff $\alpha$-divergence for $\alpha = \frac{1}{2}$, upper and lower bounds the BER \cite{bhattacharyya1946measure, kailath1967divergence}. The BC bounds are often used as motivation for algorithms in the statistical learning literature because these bounds have closed form expressions for many commonly used distributions. In addition, for small differences between the two classes, it has been shown that, in the class of Chernoff $\alpha$-divergence measures,  $\alpha= \frac{1}{2}$ (the BC divergence) results in the tightest upper bound on the probability of error \cite{hero01}. 

Beyond the bounds on the BER based on the divergence measures, a number of other bounds exist based on different functionals of the distributions. In \cite{hashlamoun1994tight}, the authors derive a new functional based on a Gaussian-Weighted sinusoid that yields tighter bounds on the BER than other popular approaches. Avi-Itzhak proposes arbitrarily tight bounds on the BER in \cite{avi1996arbitrarily}. Both of these sets of bounds are tighter than the bounds we derive here; however, these bounds cannot be estimated without at least partial knowledge of the underlying distribution. A strength of the bounds proposed in this paper is that they are empirically estimable without knowing a parametric model for the underlying distribution. 

In addition to work on bounding the Bayes error rate, recently there have been a number of attempts to bound the error rate in classification problems for the case where the training data and test data are drawn from different distributions (an area known as domain-adaptation or transfer learning in the machine learning literature). In \cite{ben2007analysis, ben2010theory}, Ben-David \emph{et al.} relate the expected error on the test data to the expected error on the training data, for the case when no labeled test data is available. In \cite{blitzer2008learning}, the authors derive new bounds for the case where a small subset of labeled data from the test distribution is available. In \cite{mansour2009domain}, Mansour \emph{et al.} generalize these bounds to the regression problem. In \cite{mansour2009multiple}, the authors present a new theoretical analysis of the multi-source domain adaptation problem based on the $\alpha$-divergence. In contrast to these models, we propose a general non-parametric bound that can be estimated without assuming an underlying model for the data and without restrictions on the hypothesis class. 

While previous bounds have proven useful in a number of applications, a drawback shared by most divergence functions (and corresponding bounds) is that they require some knowledge of the underlying distribution for their estimation. For some of the more popular divergence measures, closed form solutions are available for different distribution types \cite{fukunaga1990introduction}. More recently, a number of non-parametric methods have been introduced to estimate information theoretic quantities. Graph-based non-parametric estimators were introduced in \cite{costa2004geodesic}. Plug-in estimates of existing divergence measures that require density estimation have also been proposed  \cite{sricharan2012estimation}. More recently, estimates of the KL divergence that rely on estimates of the likelihood ratio instead of direct density estimation have been proposed \cite{nguyen2009surrogate, nguyen2010estimating}. In \cite{berisha2015Empirical} a minimal spanning tree (MST) based estimator of a different kind of $f$-divergence measure was investigated. Unlike other divergences, this $f$-divergence can be estimated directly from the data without performing density estimation. This estimator was used in \cite{berisha2015Empirical} to develop a nonparametric estimator for the Fisher information.  Whereas that paper analyzes the utility of the proposed $f$-divergence for estimation problems, this work focuses on its importance to binary classification tasks.

The rest of this paper is outlined as follows: In section II, we provide an overview of the divergence measure  and its consistent estimator. In section III, we derive bounds on the BER based on this probability distance measure and compare the tightness of the bound with Bhattacharya bound and, more generally, with the bound based on the $\alpha$-divergence. In section IV, we derive bounds on the classification error rate for the case where the training and the test data come from different distributions. In section V, we provide numerical results that confirm the validity of the bounds and describe two practical algorithms for feature learning that aim to minimize the upper bound on the error rate. Section VI contains concluding remarks and a discussion of future work.

\section{A Nonparametric Divergence Measure}
\label{sec:divergence}

For parameters $p \in (0,1)$ and $q =1-p$ consider the following divergence measure between distributions $f$ and $g$ with domain $\Reals^d$:
\begin{equation}
\label{eqn:div_HP}
D_{p}(f,g) = \frac{1}{4pq}\left [ \int \frac{(p f(\mathbf{x})- qg(\mathbf{x}))^2}{p f(\mathbf{x}) + qg(\mathbf{x})} d\mathbf{x} - (p-q)^2 \right]
\end{equation}

The divergence in (\ref{eqn:div_HP}), first introduced in \cite{berisha2015Empirical}, has the remarkable property that it can be estimated directly without estimation or plug-in of the densities $f$ and $g$ based on an extension of the Friedman-Rafsky (FR) multi-variate two sample test statistic \cite{friedman1979multivariate}. Let us consider sample realizations from $f$ and $g$, denoted by  $\mathbf{X}_f \in \Reals^{N_f \times d}$, $\mathbf{X}_g \in \Reals^{N_g \times d}$. The FR test statistic, $\mathcal{C}(\mathbf{X}_f, \mathbf{X}_g)$, is constructed by first generating a Euclidean minimal spanning tree (MST) on the concatenated data set, $\mathbf{X}_f \cup \mathbf{X}_g$, and then counting the number of edges connecting a data point from $f$ to a data point from $g$. The test assumes a unique MST for $\mathbf{X}_f \cup \mathbf{X}_g$ - therefore all inter point distances between data points must be distinct. However, this assumption is not restrictive since the MST is unique with probability one when $f$ and $g$ are Lebesgue continuous densities. In Theorem 1, we present an estimator that relies on the FR test statistic and asymptotically converges to $D_p(f,g)$. Note that this theorem combines the results of Theorem 1 and equations (3) and (4) in \cite{berisha2015Empirical}. The proof of this theorem can be found in Appendix A.

\vspace{0.3cm}
\begin{mydef}
  As $N_f \rightarrow \infty$ and $N_g \rightarrow \infty$ in a linked manner such that $\frac{N_f}{N_f+N_g}\rightarrow p$ and $\frac{N_g}{N_f+N_g}\rightarrow q$,
\[ 1 - \mathcal{C}(\mathbf{X}_f, \mathbf{X}_g)\frac{N_f + N_g}{2N_f N_g} \rightarrow D_{p}(f,g). \] almost surely.
\label{thm:HP}
\end{mydef}
\vspace{0.3cm}
  \begin{figure*}[!t]
       
        \begin{center}
\subfloat[$f(x)=g(x)$]{
        \includegraphics[width=0.225\textwidth]{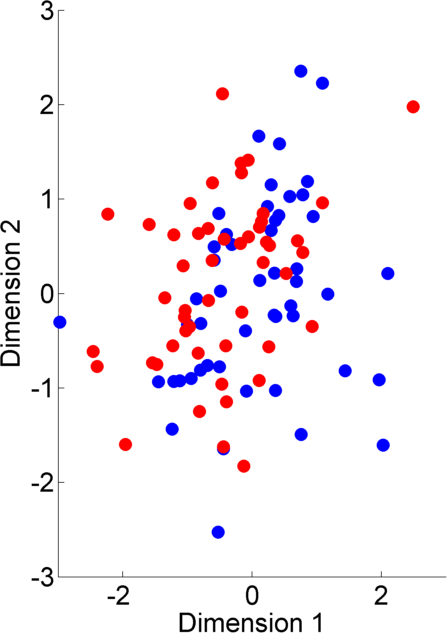}

        \includegraphics[width=0.225\textwidth]{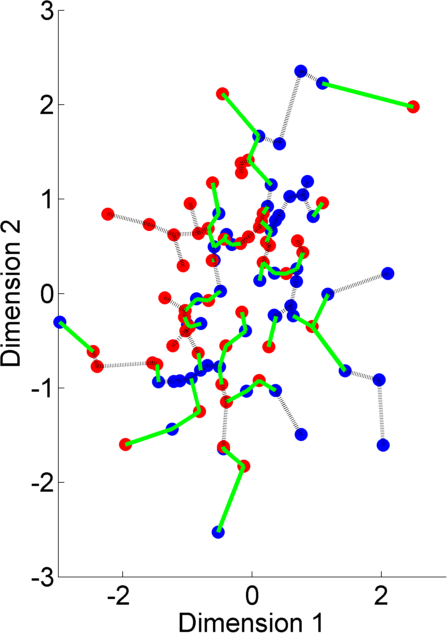}
        \label{fig:exmst1}
       }
       \subfloat[$f(x)\neq g(x)$]{
                \includegraphics[width=0.225\textwidth]{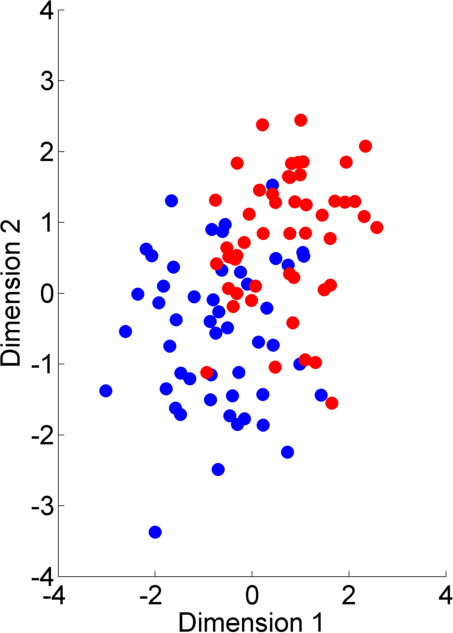}

                \includegraphics[width=0.225\textwidth]{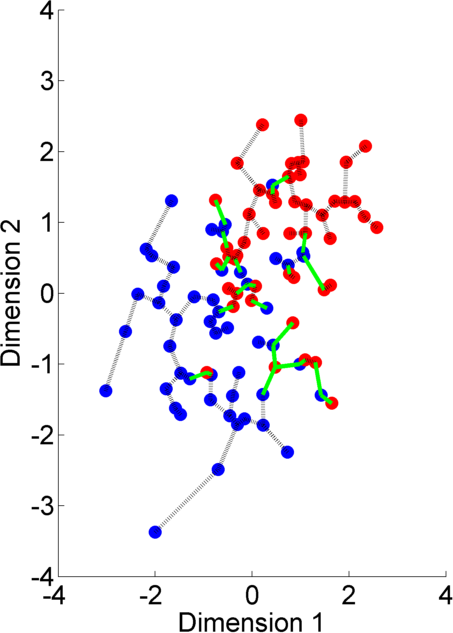}
                \label{fig:exmst2}
              }
       
       \caption{Estimation of $D_p$ for the case when (a) $f=g$ and (b) $f \neq g$.}
       \label{fig:ExMSTs}
       \end{center}
       \end{figure*} 
       
In Fig.\ref{fig:exmst1} and \ref{fig:exmst2} we show two numerical examples in order to visualize the results of Theorem \ref{thm:HP} - we plot samples from two distributions, $\mathbf{X}_f \sim f(\mathbf{x})$ and $\mathbf{X}_g \sim g(\mathbf{x})$, and evaluate the value of $\mathcal{C}(\mathbf{X}_f,\mathbf{X}_g)$. In Fig.\ref{fig:exmst1}, both data sets are drawn from the same distribution, $f(\mathbf{x}) = g(\mathbf{x}) = \mathcal{N}([0, 0]^{\mathrm{T}},\mathbf{I})$. In Fig. \ref{fig:exmst2}, we plot data drawn from $f(\mathbf{x}) = \mathcal{N}([-\frac{\sqrt{2}}{2}, -\frac{\sqrt{2}}{2}]^{\mathrm{T}},\mathbf{I})$ and $g(\mathbf{x}) = \mathcal{N}([\frac{\sqrt{2}}{2}, \frac{\sqrt{2}}{2}]^{\mathrm{T}},\mathbf{I})$. $\mathbf{I}$ is the identity matrix. For both data sets, an equal number of points are drawn, therefore $N_f = N_g = N$ and $p=q=\frac{1}{2}$. The dotted line in each figure represents the Euclidean MST associated with $\mathbf{X}_f \cup \mathbf{X}_g$. The green lines represent the edges of the MST connecting points from $f$ to points from $g$, $\mathcal{C}(\mathbf{X}_f,\mathbf{X}_g)$. We can use this to estimate $D_p(f,g)$ using the results of Theorem \ref{thm:HP}. It is clear from the figures that this value is much smaller for overlapping distributions (Fig. \ref{fig:exmst1}) than for separable distributions (Fig. \ref{fig:exmst2}). Indeed, as Theorem \ref{thm:HP} suggests, in the limit, this statistic converges to the integral used in the divergence measure in (\ref{eqn:div_HP}).  

In the ensuing sections we outline some important properties of this divergence measure and develop new bounds for classification using this distance function between distributions.

\subsection{Properties of $D_p$}

The divergence measure in (\ref{eqn:div_HP}) exhibits several properties that make it useful for statistical analysis. It is relatively straightforward to show that the following three properties are satisfied. 

\begin{enumerate}
\item $0 \leq D_{p} \leq 1$
\item $D_{p} = 0 \iff f(\mathbf{x})=g(\mathbf{x})$.
\item $D_{p}(f,g)=D_{q}(g,f)$
\end{enumerate}

The lower bound in the first property follows from the fact that when $f=g$ and $p=q$, the minimum value of $D_{p}$ is 0. To show that the divergence measure is upper bounded by 1, we first note that
\begin{align}
\int \frac{(p f(\mathbf{x})- qg(\mathbf{x}))^2}{p f(\mathbf{x}) + qg(\mathbf{x})} d\mathbf{x} = 1 - 4pq A_p(f,g), 
\label{eqn:defineA}
\end{align}
where 
\begin{equation*}
A_p(f,g) = \int \frac{ f(\mathbf{x}) g(\mathbf{x})}{p f(\mathbf{x}) + qg(\mathbf{x})} d\mathbf{x}.
\end{equation*}
The function $A_p(f,g)$ attains its minimum value of 0, when $f$ and $g$ have no overlapping support (since $f(\mathbf{x}) > 0$ and $g(\mathbf{x}) > 0$ for all $\mathbf{x}$); therefore $D_p = \frac{1}{4pq} [1 - (p-q)^2] = 1$. The second property is closely related to the first: the minimum value $D_{p} = 0$ only when $f=g$ and $p=q$. The third property follows from commutativity. 

The divergence measure in (\ref{eqn:div_HP}) belongs to the class of $f$-divergences. Every $f$-divergence can be expressed as an average of the ratio of two distributions, weighted by some function $\phi(t)$: $D_{\phi}(f,g) = \int \phi(\frac{f(\mathbf{x})}{g(\mathbf{x})}) g(\mathbf{x}) d\mathbf{x}$. For $D_{p}(f,g)$, the corresponding function $\phi(t)$ is,
\begin{equation}
\phi(t) = \frac{1}{4 p q} \Bigg [ \frac{(p t- q)^2}{p t+ q} - (2 p - 1)^2 \Bigg].
\end{equation}
Furthermore, $\phi(t)$ is defined for all $t>0$, is convex - $\phi''(t) = \frac{2pq}{(pt+q)^3} > 0$, and $\phi(1) = 0$. This is consistent with the requirements of the definition of an $f$-divergence \cite{csiszar04}. Indeed, for the special case of $\alpha=\frac{1}{2}$, the divergence in (\ref{eqn:div_HP}) becomes the symmetric $\chi^2$ $f$-divergence in \cite{Cha07} and is similar to the Rukhin $f$-divergence in \cite{Ruhkin94}.


\section{Bounds on Bayes Classification Error}
\label{sec:BER}

In this section, we show how $D_p$ in (\ref{eqn:div_HP}) can be used to bound the Bayes error rate (BER) for binary classification. Further, we show that, under certain conditions, this bound is tighter than the well-known Bhattacharya bound commonly used in the machine learning literature and can be empirically estimated from data.

Before deriving the error bounds, for notation convenience, we introduce a slightly modified version of the divergence measure in (\ref{eqn:div_HP}), 
\begin{align}
\label{eq:DHPh}
\tilde{D}_{p}(f,g) & =1-4pq\int \frac{ f(\mathbf{x}) g(\mathbf{x})}{p f(\mathbf{x}) + qg(\mathbf{x})} d\mathbf{x} \\
 & = \int \frac{(pf(\mathbf{x}) - qg(\mathbf{x}))^2}{pf(\mathbf{x}) + qg(\mathbf{x})} d\mathbf{x}. \nonumber
\end{align}

It is easy to see that $D_{p} = \frac{\tilde{D}_{p}}{4pq} - \frac{(p-q)^2}{4pq}$ and when $p=q=0.5$, $D_{p} = \tilde{D}_{p}$. While this function no longer satisfies $\tilde{D}_{p}(f,g) = 0$, for $f=g$, and therefore is no longer a valid divergence measure, it greatly simplifies the notation of the ensuing error bounds. As with $D_p$, w can estimate this quantity using the FR test statistic since, under the same conditions as those in Theorem 1,
\begin{equation}
1 -2\frac{\mathcal{C}(\mathbf{X}_f, \mathbf{X}_g)}{N_f + N_g} \rightarrow \tilde{D}_{p}(f,g).
\end{equation}

Given a binary classification problem with binary labels  $y \in \{0, 1\}$ and $\mathbf{x}$ drawn from $f_{\mathrm{S}}(\mathbf{x})$, we denote the conditional distributions for both classes as $f_0(\mathbf{x}) = f_\mathrm{S}(\mathbf{x} | y=0)$ and $f_1(\mathbf{x}) = f_\mathrm{S}(\mathbf{x} | y=1)$. We draw samples from these distributions with probability $p$ and $q=1-p$, respectively, and formulate two data matrices denoted by $\mathbf{X}_0 \in \Reals^{N_0 \times d}$ and $\mathbf{X}_1 \in \Reals^{N_1 \times d}$. The Bayes error rate associated with this problem is given in (\ref{eq:BER}). In Theorem \ref{thm:BER} below, we show that we can bound this error from above and below using the divergence measure introduced in the previous section. The proof of this theorem can be found in Appendix B.

%

\vspace{0.3cm}
\begin{mydef} For two distributions, $f_0({\mathbf{x}})$ and $f_1({\mathbf{x}})$, with prior probabilities $p$ and $q$ respectively, the Bayes error rate, $\epsilon^{\mathrm{Bayes}}$, is bounded above and below as follows:
\begin{equation*}
\frac{1}{2} - \frac{1}{2}\sqrt{\tilde{D}_{p}(f_0,f_1)} \leq \epsilon^{\mathrm{Bayes}} \leq \frac{1}{2} - \frac{1}{2}\tilde{D}_{p}(f_0,f_1).
\end{equation*}
\label{thm:BER}
\end{mydef}

Combining the results from Theorem \ref{thm:HP} with the results of Theorem \ref{thm:BER}, we see that we can approximate the upper and lower bounds on the BER from the data matrices $\mathbf{X}_0$ and $\mathbf{X}_1$ as
 \begin{equation*}
 \frac{1}{2} - \frac{1}{2}\sqrt{\tilde{D}_{p}(f_0,f_1)} \approx \frac{1}{2} - \frac{1}{2}\sqrt{1-2 \frac{\mathcal{C}(\mathbf{X}_0,\mathbf{X}_1)}{N_0 + N_1}},
 \end{equation*}
 and
 \begin{equation*}
 \frac{1}{2} - \frac{1}{2} \tilde{D}_{p}(f_0,f_1) \approx \ \frac{\mathcal{C}(\mathbf{X}_0,\mathbf{X}_1)}{N_0 + N_1}.
 \end{equation*}
The derived bound is tight for the case  $p=q=\frac{1}{2}$. For $f_0(\mathbf{x}) = f_1(\mathbf{x})$, the BER is 0.5. Under these conditions, $\tilde{D}_{p}(\mathbf{x}) = 0$, and both the upper and lower bound in Theorem \ref{thm:BER} go to 0.5. For the case where $f_0(\mathbf{x})$ and $f_1(\mathbf{x})$ are completely separable, the BER is 0, $\tilde{D}_{p}(\mathbf{x}) = 1$, and both the upper and lower bound go to 0. 

 \subsection{Relationship to the Chernoff Information Bound}

Here we compare the tightness of the bounds on the Bayes error rate based on $D_p$ to the bounds based on the Chernoff information function (CIF) \cite{hero01}, defined as 
\[ 
I_{\alpha}(f_0, f_1) = \int p^{\alpha}f_0^{\alpha}(\mathbf{x}) q^{1-\alpha}f_1^{1-\alpha}(\mathbf{x}) d\mathbf{x}. 
\] 
In Theorem \ref{thm:HPChernoff}, we derive an important relationship between the affinity measure, $A_p(f_0, f_1)$, and a scaled version of the CIF. The proof of this theorem can be found in Appendix C.

\vspace{0.3cm}
\begin{mydef} The affinity measure, $A_p(f_0, f_1)$, is a lower bound for a scaled version of the Chernoff information function:
\begin{equation*}
A_p(f_0,f_1)\leq  \int f_0^q (\mathbf{x})f_1^p(\mathbf{x}) d\mathbf{x}.
\end{equation*}
\label{thm:HPChernoff}
\end{mydef}
\vspace{0.1cm}

It is important to note that the second term in Theorem \ref{thm:HPChernoff} is exactly equal to the CIF for $\alpha = p = q = 1/2$. For this special case, the Chernoff bound reduces to the Bhattacharyya (BC) bound, a widely-used bound on the Bayes error in machine learning that has been used to motivate and develop new algorithms \cite{kailath1967divergence, saon01, xuan06}. The popularity of the BC bound is mainly due to the the fact that closed form expressions for the bound exist for many of the commonly used distributions. Let us define the Bhattacharya coefficient as:

\begin{equation}
BC(f_0,f_1) = 2 \int \sqrt{pq f_0(\mathbf{x}) f_1(\mathbf{x})} d\mathbf{x}.
\end{equation}
The well-known Bhattacharya bound on the BER is given by
\begin{equation} \label{eqn:bcbound}
\frac{1}{2}-\frac{1}{2}\sqrt{1-BC^2(f,g)} \leq \epsilon^{\mathrm{Bayes}} \leq \frac{1}{2} BC(f,g).
\end{equation}

In Theorem \ref{thm:tightness} below, we show that, for equiprobable classes, the $D_p$ bound provides tighter upper and lower bounds on the BER when compared to the bound based on the BC coefficient under all separability conditions. The proof of this theorem can be found in Appendix D.
 
\vspace{0.3cm}
\begin{mydef} For $p = q = \frac{1}{2}$, the $D_p$ upper and lower bounds on the Bayes error rate are tighter than the Bhattacharyya bounds:
\begin{equation}
\begin{aligned}
 \frac{1}{2}-\frac{1}{2}\sqrt{1-BC^2(f_0,f_1)} \leq & \frac{1}{2} - \frac{1}{2}\sqrt{\tilde{D}_{\frac{1}{2}}(f_0,f_1)}  \\   
\leq \epsilon^{\mathrm{Bayes}} \leq   \frac{1}{2} - & \frac{1}{2}\tilde{D}_{\frac{1}{2}}(f_0,f_1) \leq \frac{1}{2}BC(f_0,f_1). \nonumber
 \end{aligned}
 \label{eq:thm3}
 \end{equation}
 \label{thm:tightness}
\end{mydef}
\vspace{0.1cm}

Using asymptotic analysis of the Chernoff exponent, for small differences between the two classes, it was shown that $\alpha=\frac{1}{2}$ results in the tightest bound on the probability of error - this corresponds to the bound in (\ref{eqn:bcbound}) \cite{hero01}. Using a variant of this analysis, we derive a local representation of the CIF and relate it to the divergence measure proposed here. In particular, if we let 
\begin{align*}
pf_0(\mathbf{x}) & = \frac{1}{2}(pf_0(\mathbf{x}) + qf_1(\mathbf{x})) + \frac{1}{2}(pf_0(\mathbf{x}) - qf_1(\mathbf{x})) \\ 
& =f_{\frac{1}{2}}    (\mathbf{x})   (1+\frac{1}{2}\Delta_{\mathbf{x}}), 
\end{align*} 
where $f_{\frac{1}{2}} (\mathbf{x}) =\frac{1}{2}(pf_0(\mathbf{x}) + qf_1(\mathbf{x})) $ and $\Delta_{\mathbf{x}} = (pf_0(\mathbf{x}) - qf_1(\mathbf{x}))/f_{\frac{1}{2}} (\mathbf{x})$. Similarly, 
\begin{equation*}
qf_1(\mathbf{x}) = f_{\frac{1}{2}}(\mathbf{x})(1-\frac{1}{2}\Delta_{\mathbf{x}}). 
\end{equation*}
As in \cite{hero01}, after a Taylor series expansion around $p^{\alpha}f_0^{\alpha}(\mathbf{x})$ and $q^{1-\alpha}f_1^{1-\alpha}(\mathbf{x})$, the Chernoff information function can be expressed as (see proof of Proposition 5 in \cite{hero01}):
\begin{align} \nonumber
I_{\alpha}(f_0, f_1)  & = \int f_{\frac{1}{2}}(\mathbf{x})  \bigg [ 1 - (2\alpha-1)\frac{\Delta_{\mathbf{x}}}{2} \\ \nonumber 
 & \ \ \ \ \  -  \alpha(1-\alpha) \left( \frac{\Delta_{\mathbf{x}}}{2} \right)^2 + o(\Delta_{\mathbf{x}}^3) \bigg ] d\mathbf{x} \\ \nonumber
& = \int f_{\frac{1}{2}}(\mathbf{x}) dx - (2\alpha - 1) \int f_{\frac{1}{2}}(\mathbf{x}) \frac{\Delta_{\mathbf{x}}}{2} d\mathbf{x} \\ \nonumber
& \ \ \ \ \   -\alpha(1-\alpha) \int f_{\frac{1}{2}}(\mathbf{x}) \left( \frac{\Delta_{\mathbf{x}}}{2} \right ) ^2 + o(\Delta^2)\\ \nonumber
& = \frac{1}{2} - (2\alpha-1)(2p-1)/2  \\ \nonumber
& \ \ \ \ \  -  \frac{\alpha(1-\alpha)}{2}\int \frac{(pf_0(\mathbf{x}) - qf_1(\mathbf{x}))^2}{pf_0(\mathbf{x}) + qf_1(\mathbf{x})} d\mathbf{x} + o(\Delta^2) \\ \nonumber
& = (p+\alpha) - 2\alpha p - \frac{\alpha(1-\alpha)}{2} \tilde{D}_{p}(f_0,f_1) + o(\Delta^2) \nonumber
\end{align}

The local equivalence of $D_p$ and $I_\alpha$ is not surprising since all $f$-divergences are locally equivalent (they induce the same Riemann-Fisher metric on the manifold of densities) \cite{csiszar04}. This useful property allows us to estimate the CIF for small differences between $f_0$ and $f_1$ using the MST procedure in Section \ref{sec:divergence}. Further, we can express the BER in terms of the CIF: 
\begin{equation}
\epsilon^{\mathrm{Bayes}} \leq I_{\alpha} \approx (p+\alpha) - 2\alpha p - \frac{\alpha(1-\alpha)}{2} \tilde{D}_{p}(f_0,f_1). \nonumber
\end{equation}
For $p = q = \frac{1}{2}$, this bound reduces to $\epsilon^{\mathrm{Bayes}} \leq \frac{1}{2} - \frac{\alpha(1-\alpha)}{2} \tilde{D}_{\frac{1}{2}}(f_0,f_1)$. This is very similar to the upper bound in Theorem \ref{thm:BER}, differing only in the scale of the second term. Further, it is easy to see from this that the bound in Theorem \ref{thm:BER} is tighter than the Chernoff bound since $\frac{\alpha(1-\alpha)}{2} < \frac{1}{2}$ for all $\alpha$. This is not surprising since, locally, $\alpha=0.5$ yields the tightest bounds on the BER \cite{hero01}. This corresponds to the BC bound in (\ref{eqn:bcbound}) and we have already shown that new bound is tighter than the BC bound in Theorem \ref{thm:tightness}. This analysis further confirms that result.

%
%
  
\begin{figure}
      \centerline{\includegraphics[width=.5\textwidth]{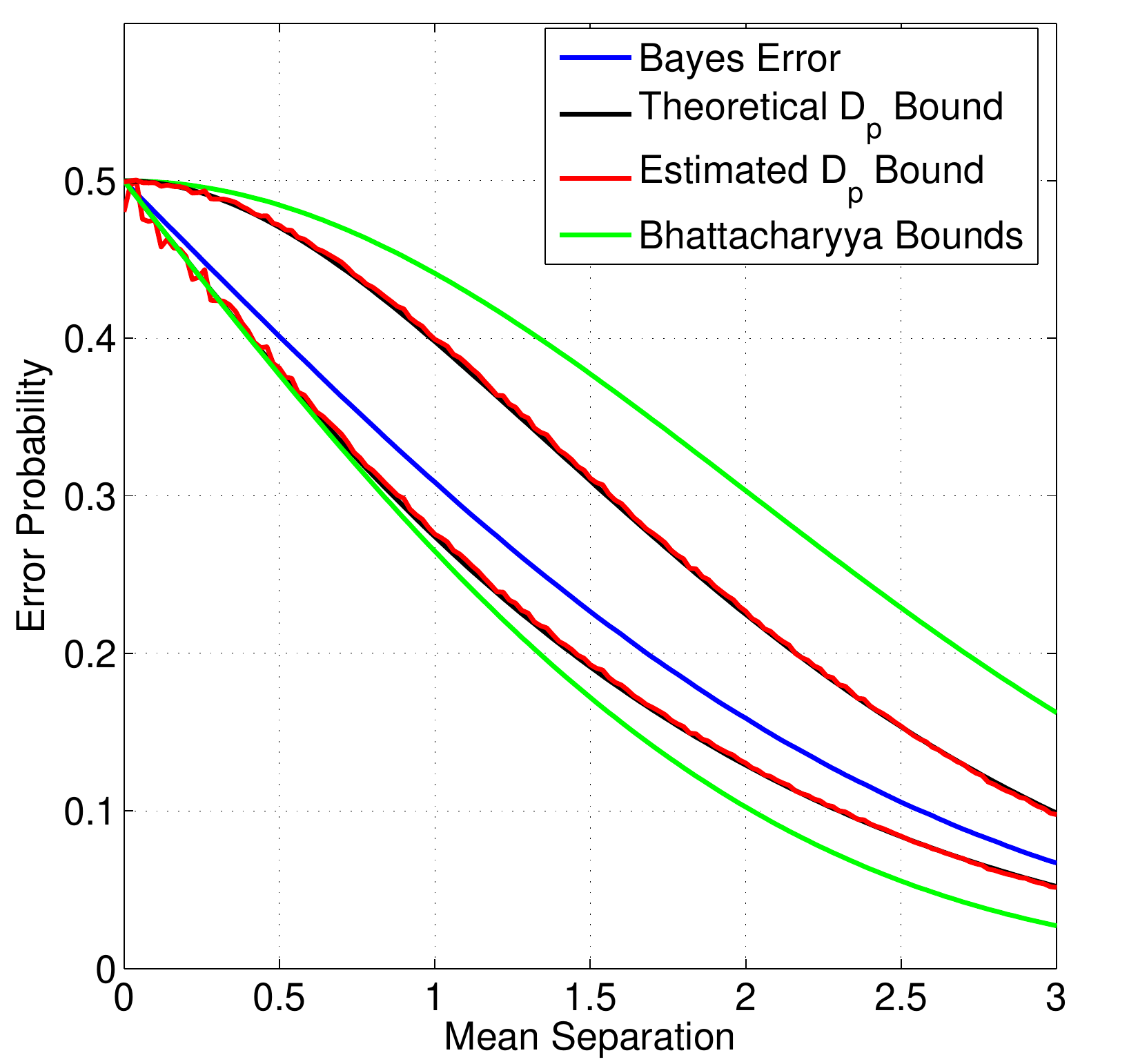}}
      \caption{The $D_p$ and BC bounds on the Bayes error rate for a bivariate Gaussian example.}
        \label{fig:BayesBound}
\end{figure}

In addition to providing tighter bounds on the BER we can estimate the new $D_p$ bound without ever explicitly computing density estimates. We provide a numerical example for comparison. We consider two data samples from two classes, each of which comes from a normally distributed bivariate distribution with varying mean and spherical unit variance. The separation in means between the two class distributions is increased incrementally across 150 trials. The two distributions completely overlap initially, and are almost entirely separated by the final trial. In each trial we calculate the BER analytically using (\ref{eq:BER}), as well as the upper and lower bounds introduced in Theorem \ref{thm:BER}. We calculate the bounds both analytically (through numerical integration) and empirically (using the results from Theorem \ref{thm:HP}). In order to demonstrate the tightness of this bound we also plot it against the upper and lower Bhattacharyya error bounds for Gaussian data (the closed form expression of the bound for Gaussian data is known) \cite{kailath1967divergence}. Figure \ref{fig:BayesBound} displays the true BER along with both error bounds as a function of the Euclidean separation between the means of two bivariate normal distributions of unit variance. We see in this plot that the proposed error bounds are noticeably tighter than the Bhattacharyya error bounds and are well correlated with the true BER. Although the analytically calculated $D_p$ bound never crosses the BC bound, the empirically estimated $D_p$ bound crosses the BC bound for small values of the mean separation. This is due to the variance of the estimator. It is important to note that the estimator used here {\em asymptotically} converges to the $D_p$ divergence; however this result doesn't necessarily extend to finite data. In fact, for any fixed estimator, there exists a distribution for $X$ and $y$ such that the error converges arbitrarily slowly \cite{antos99}.

%

\section{Bounds on the Domain Adaptation Error}
\label{sec:DAbounds}
In this section, we consider a cross-domain binary classification problem and show how the $D_p$ distance can be used to bound the error rate in this setting also. Let us define data from two domains, the source (training) and the target (testing) domain and the corresponding labeling functions for each domain  $y_{\mathrm{S}}(\mathbf{x}), y_{\mathrm{T}}(\mathbf{x}) \in \{0, 1\}$ that yields the true class label of a given data point $\mathbf{x}$. The source domain, denoted by the pair $(\mathbf{X}_{\mathrm{S}}, y_{\mathrm{S}})$, represents the data used to train the machine learning algorithm and the data $(\mathbf{X}_{\mathrm{T}}, y_{\mathrm{T}})$ represents the data the algorithm will encounter once deployed. Let us further define the conditional distributions $f_{\mathrm{S},0}(\mathbf{x}) = f_{\mathrm{S}}(\mathbf{x} | y_s(\mathbf{x}) = 0)$ and $f_{\mathrm{S},1}(\mathbf{x}) = f_{\mathrm{S}}(\mathbf{x} | y_s(\mathbf{x}) = 1)$. The rows of the source and target data are drawn from $f_{\mathrm{S}}(\mathbf{x})$ and $f_{\mathrm{T}}(\mathbf{x})$. The risk, or the probability that the decision, $\mathit{h}$, disagrees with the true label is defined as 
\begin{equation}
\epsilon_{\mathrm{S}} (h, y_{\mathrm{S}}) = \mathbf{E}_{f_{\mathrm{S}}(x)}   [|h(\mathbf{x}) - y_{\mathrm{S}} |],
\end{equation}
for the source data. It is similarly defined for the target data. In Theorem \ref{thm:DAbound}, we identify a relationship between the error rates on the source and target data.  The proof of this theorem can be found in Appendix E.

\vspace{0.3cm}
\begin{mydef}
Given a hypothesis, $h$, the target error, $\epsilon_{\mathrm{T}}(h, y_{\mathrm{T}})$, can be bounded by the error on the source data, $\epsilon_{\mathrm{S}} (h, y_{\mathrm{S}}) $, the difference between labels, and a distance measure between source and target distributions as follows:
\begin{align}
\epsilon_{\mathrm{T}} (h, y_{\mathrm{T}}) \leq & \epsilon_{\mathrm{S}} (h, y_{\mathrm{S}}) +   \mathbf{E}_{f_{\mathrm{S}}(x)}   [|y_{\mathrm{S}} - y_{\mathrm{T}} |] \\ \nonumber + &  2\sqrt{\tilde{D}_{\frac{1}{2}} ( f_{\mathrm{S}},  f_{\mathrm{T}} )},
\end{align}
where $\tilde{D}_{\frac{1}{2}} ( f_{\mathrm{S}},  f_{\mathrm{T}} )$ assumes equiprobable data from the source and target distributions.
\label{thm:DAbound}
\end{mydef}
\vspace{0.3cm}

The bound in Theorem \ref{thm:DAbound} depends on three terms: the error on the source data, the expected difference in the labeling functions across the two domains, and a measure of the distance between source and target distributions ($D_p$ distance). We expect that the selected training algorithm will seek to minimize the first term; the second term characterizes the difference between labeling functions in the source and target domains; the third term is of particular interest to us - it provides a means of bounding the error on the {\em target} data as a function of the distance between source and target distributions. 

In the {\em covariate shift} scenario, we assume that there exists no difference between labeling functions (e.g. $y_{\mathrm{S}} (\mathbf{x}) = y_{\mathrm{T}} (\mathbf{x})$) and only the distributions between the source and target data change \cite{ben2010theory}. Under this assumption, the bound in Theorem \ref{thm:DAbound} reduces to
\begin{equation}
\epsilon_{\mathrm{T}} (h, y_{\mathrm{T}}) \leq \epsilon_{\mathrm{S}} (h, y_{\mathrm{S}}) +  2\sqrt{\tilde{D}_{\frac{1}{2}} ( f_{\mathrm{S}},  f_{\mathrm{T}} )}.
 \label{eq:BD}
\end{equation}
                                  
Furthermore, if we assume that the decision rule $h$ attains the Bayes error rate, $\epsilon^{\mathrm{Bayes}}$, on the source domain, we can use the results from Theorem \ref{thm:BER} to rewrite the bound in Theorem \ref{thm:DAbound} using only the $D_p$ distance:
                                  
\begin{equation}
\epsilon_{\mathrm{T}}  \leq \frac{1}{2} - \frac{1}{2}\tilde{D}_{p} ( f_{\mathrm{S},0},  f_{\mathrm{S},1} ) +  2\sqrt{\tilde{D}_{\frac{1}{2}} ( f_{\mathrm{S}},  f_{\mathrm{T}} )}.
\label{eq:et1}
\end{equation}
If we denote the training data matrices by $\mathbf{X}_{\mathrm{S},0} \sim  f_{\mathrm{S},0}$ and $\mathbf{X}_{\mathrm{S},1}\sim  f_{\mathrm{S},1}$, then we can estimate this upper bound using the FR test statistic by
\begin{equation}
\frac{\mathcal{C}(\mathbf{X}_{\mathrm{S},0},\mathbf{X}_{\mathrm{S},1})}{N_{\mathrm{S},0} + N_{\mathrm{S},1}} + 2\sqrt{1-2\frac{\mathcal{C}(\mathbf{X}_{\mathrm{S}},\mathbf{X}_{\mathrm{T}})}{N_{\mathrm{S}} + N_{\mathrm{T}}}}.
\label{eq:et2}
\end{equation}
         
             The result shown in (\ref{eq:et2}) represents an upper bound on the target domain error that can be computed without access to any labels in this domain. This bound provides interesting insight on the importance of invariant representations for classification. The target error is bounded by the sum of the affinity between class distributions in the source domain and the square root of the $D_p$-distance between domains. Because of the square root and the multiplicative factor, it is clear that the second term in (\ref{eq:et2}) is weighted much more heavily. This stresses the importance of invariant representations in classification. In other words, the bound provides a means of quantifying the relative importance of selecting features that are invariant across domains versus features that provide good separation separation between classes in the source domain. 

\section{Numerical Results and Practical Algorithms}
\label{sec:Practical Algorithms}

Here, we describe a number of numerical experiments that evaluate the bounds in a classification setting. In the first experiment, we evaluate the tightness of the bound on the Bayes error rate in higher dimensions by comparing against two other bounds for an example where the Bayes error rate is known in closed form. In the second and third experiments, we develop new criteria for feature selection based on the derived bounds and compare the probability of correct classification against competing alternatives.

\subsection{Bounding the Bayes Error Rate}
Consider the two data sets $\mathcal{D}_1$ and $\mathcal{D}_2$ in Table \ref{tbl:BERDataSets}, each consisting of data from two 8 dimensional Gaussian distributions. In \cite{fukunaga90} Fukunaga computed the true Bayes error rate analytically for both of these data sets. Here we compare three different bounds on this error for both datasets -  the $D_p$-based bound, the Mahalanobis bound, and the BC bound. We use the closed-form version of the BC and Mahalanobis bound for Gaussian data \cite{fukunaga90}. Furthermore, we assume perfect knowledge of the parameters for these two bounds ($\sigma$ and $\mu$). As a result, this is the best possible case for both of these bounds - the data matches the model and no estimation of the parameters is required.

\renewcommand{\tabcolsep}{5pt}
\begin{table}
\caption{Parameters for 2 8-dimensional Gaussian data sets for which the Bayes error rate is known (from \cite{fukunaga90})}
\begin{tabular}{c  c | c  c  c  c  c  c  c  c  }
\hline
 \multirow{4}{*}{$\mathcal{D}_1$} & $\mu_1$ & 0& 0& 0& 0& 0& 0& 0& 0 \\
                              & $\sigma_1$ & 1& 1& 1& 1& 1& 1& 1& 1 \\
                              & $\mu_2$ & 2.56 & 0& 0& 0& 0& 0& 0& 0 \\
                              & $\sigma_2$ & 1& 1& 1& 1& 1& 1& 1& 1 \\ \hline
     \multirow{4}{*}{$\mathcal{D}_2$} & $\mu_1$ & 0& 0& 0& 0& 0& 0& 0& 0 \\
                               & $\sigma_1$ & 1& 1& 1& 1& 1& 1& 1& 1 \\
                                & $\mu_2$ & 3.86 & 3.10 & 0.84 & 0.84 & 1.64 & 1.08 & 0.26& 0.01 \\
                               & $\sigma_2$ & 8.41 & 12.06& 0.12 & 0.22 & 1.49 & 1.77& 0.35 & 2.73 \\ \hline
\end{tabular}
\label{tbl:BERDataSets}
\end{table}

For both data sets $\mathcal{D}_1$ and $\mathcal{D}_2$, we evaluate the $D_p$-based upper bound between the two distributions using the graph-based method outlined in Section \ref{sec:divergence} for three different sample sizes (100 samples, 500 samples, and 1000 samples - 50 Monte Carlo simulations each). We compare the $D_p$ bound (computed from empirical data without assuming any parametric model of the data distribution) with the Bhattacharyya bound and the Mahalanobis bound. For both data sets, the average $D_p$-based bound is closer to the true error rate, regardless of the sample size. Again, it is important to stress that this is the best case scenario for the competing bounds since there exists a closed form expression for both bounds for Gaussian data and we assume perfect knowledge of the distribution parameters. Regardless, the empirically-estimated $D_p$ bound is still tighter.

\begin{table}
\caption{Comparing upper bounds on the Bayes error rate for the multivariate Gaussians defined in Table \ref{tbl:BERDataSets}.}
\centering
\begin{tabular}{l c c }
\hline\hline
 & Data 1 & Data 2 \\ [0.5ex] 
\hline
Actual Bayes Error & 10\% & 1.90\% \\
Mahalanobis Bound &18.95\%& 14.13\% \\
Bhattacharyya Bound &22.04\% & 4.74\% \\
$D_p$ Bound (100 points) & \bf{18.23\%} $\pm$ \bf{3.32\%} &  \bf{4.10\%} $\pm$ \bf{1.10\%}  \\
$D_p$ Bound (500 points) & \bf{16.88\%} $\pm$ \bf{1.51\%} & \bf{2.17\%} $\pm$ \bf{0.42\%}  \\
$D_p$ Bound (1000 points) & \bf{16.46\%} $\pm$ \bf{1.14\%}& \bf{1.94\%} $\pm$ \bf{0.29\%}  \\
\hline
\end{tabular}
\label{table:nonlin}
\end{table}

\subsection{Feature Selection using $D_p$-distance}

In machine learning, feature selection algorithms are often used to reduce model complexity and prevent over-fitting \cite{liu2007computational}. In many scenarios, feature selection can actually improve model performance since the reduced dimensionality leads to a much more densely populated hypothesis space. This prevents the model from learning irrelevant patterns in the training data that aren't pertinent for a given task and will not generalize to new datasets. This problem is exacerbated in domain adaptation problems where the separation in domains makes misleading patterns in the training data especially problematic. We use the bounds defined in Theorems \ref{thm:BER} and \ref{thm:DAbound} to develop new feature selection criteria that aim to directly minimize the BER bound. We consider two different scenarios: (1) one where the training data and the test data come from the same distribution and (2) another where the training data and the test data come from different distributions. For both scenarios, we seek to identify the subset of features, $\Omega$, that will minimize the ``worst-case" error. For scenario 1, this results in minimizing the upper bound in Theorem 2:

\begin{equation}
\varPhi(\Omega)  =  \frac{\mathcal{C}(\mathbf{X}_1(\Omega),\mathbf{X}_2(\Omega))}{N_1+N_2},
\label{eq:crit1}
\end{equation}
and, for scenario 2, we minimize the DA bound defined in Theorem \ref{thm:DAbound}:

\begin{equation}
\begin{aligned}
\varPhi(\Omega) &=  \frac{\mathcal{C}(\mathbf{X}_{\mathrm{S},0}(\Omega ),\mathbf{X}_{\mathrm{S},1}(\Omega))}{N_{\mathrm{S},0}+N_{\mathrm{S},1}}  \\
&+2\sqrt{1-2\frac{\mathcal{C}(\mathbf{X}_{\mathrm{S}}(\Omega),\mathbf{X}_{\mathrm{T}}(\Omega))}{N_{\mathrm{S}}+N_{\mathrm{T}}}}\end{aligned}.
\label{eq:crit2}
\end{equation}

We integrate the optimization criteria into a forward selection search algorithm in Alg. \ref{alg:DAFR feature selection algorithm}. In this algorithm, we use a parameter $\alpha$ to determine whether or not the algorithm should account for the separation between domains. For traditional machine learning problems $\alpha$ should be set to $0$. For domain adaptation problems, $\alpha$ is set to $1$ to minimize the error upper bound, or tuned based on the importance of minimizing the separation between domains. We set $\alpha$ to $1$ for all DA experiments reported in this paper - this corresponds directly to the bound in Theorem \ref{thm:DAbound}. 

\begin{algorithm} [t]
\caption{Forward selection algorithm using $D_p$-distance}
\label{alg:DAFR feature selection algorithm}
\begin{algorithmic}
\State \bf{Input:} \rm{Feature data from two different classes in the source} 
\State \, \, \, \, \, \, \, \rm{domain and unlabelled data from the target}   
\State \, \, \, \, \, \, \, \rm{domain:}$\mathbf{X}_{\mathrm{S},0}$,  $\mathbf{X}_{\mathrm{S},1}$,  $\mathbf{X}_{\mathrm{T}}$, $\alpha$
\State \bf{Output:} \rm{Top} $k$ \rm{features that minimize $\varPhi$ :
\State \, \, \, \, \, \, \, \, \, $\Omega$}
\State\bf{Define:} \,  $\Omega = \emptyset$
\State \, \, \, \, \, \, \, $F = {1 \dots M}$
\State \, \, \, \, \, \, \, $\mathbf{X}_{\mathrm{S}} = \mathbf{X}_{\mathrm{S},0} \cup \mathbf{X}_{\mathrm{S},1}$
\For {$j \in {1 \dots k}$}
\State $\varPhi = \emptyset$
\For {$F_i \in F \setminus \Omega$}
\State $
\begin{aligned}
\varPhi(F_i)  &= \frac{\mathcal{C}(\mathbf{X}_{\mathrm{S},0}(\Omega \cup F_i),\mathbf{X}_{\mathrm{S},1}(\Omega \cup F_i))}{N_{\mathrm{S},0}+N_{\mathrm{S},1}} \\ &+2 \alpha \sqrt{1-2 \frac{\mathcal{C}(\mathbf{X}_{\mathrm{S}}(\Omega \cup F_i),\mathbf{X}_{\mathrm{T}}(\Omega \cup F_i))}{N_{\mathrm{S}}+N_{\mathrm{T}}}}
\end{aligned} $
\EndFor
\State $\Omega = \Omega \cup \{\underset{F_i}{\text{argmin }}  \varPhi(F_i)\}$
\EndFor
\end{algorithmic}
\end{algorithm}

We empirically evaluate the feature selection algorithm on a pathological speech database recorded from patients with neurogenic disorders. In particular, we consider the problem of classifying between healthy and dysarthric speech. Dysarthria is a motor speech disorder resulting from an underlying neurological injury. We make use of data collected in the Motor Speech Disorders Laboratory at Arizona State University, consisting of 34 dysarthric speakers and 13 healthy speakers (H). The dysarthria speakers included: 12 speakers with ataxic dysarthria, secondary to cerebellar degeneration (A), 10 mixed flaccid-spastic dysarthria, secondary to amyotrophic lateral sclerosis (ALS), 8 speakers with hypokinetic dysarthria secondary to Parkinson's Disease (PD), and 4 speakers with hyperkinetic dysarthria secondary to Huntington's disease (HD). Each patient provided speech samples, including a reading passage, phrases, and sentences. The speech database consists of approximately 10 minutes of recorded material per speaker. These speech samples were taken from the larger pathological speech database described in \cite{Liss14}.

The recordings from each speaker were split into individual sentences by hand and features were extracted at the sentence level. Three different feature sets were used: envelope modulation spectrum (EMS) features, long-term average spectrum (LTAS) features, and ITU-T P.563 features. EMS is a representation of the slow amplitude modulations in a signal and captures aspects of the speech signal related to rhythm. The LTAS features capture atypical average spectral information in the signal. The P.563 features measure atypical and unnatural voice and articulatory quality. For a more detailed discussion of these features, we refer the readers to \cite{berisha2014modeling}.

\begin{figure}
  \centering
  \includegraphics[width=0.45\textwidth]{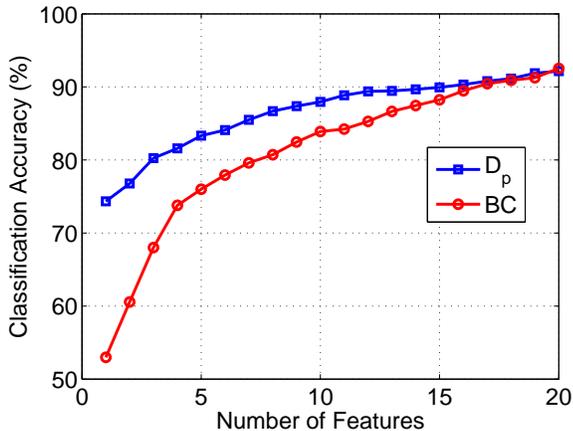}
  \caption{Average classification accuracy using reduced feature sets}
  \label{fig:ACCdys}
\end{figure}

In our first experiment we evaluate the FS algorithm based on the criteria in (\ref{eq:crit1}). We consider the problem of discriminating between healthy and dysarthric speech based on the features discussed above. For this experiment we form both the training and test sets by randomly drawing 300 dysarthric speech samples and 300 healthy speech samples for each set, ensuring that there is no overlap between training and test data. Using the FS algorithm in Alg. \ref{alg:DAFR feature selection algorithm}, we use the training data to find the top 20 features that maximize the separability between the two groups. We compare this feature selection algorithm against one based on maximizing the Bhattacharyya distance between classes. Using the feature subsets chosen by the two algorithms, we build support vector machine (SVM) classifiers on the training data and evaluate their accuracy on the test data. This experiment is repeated ten times using different randomly generated training and test sets, and the average accuracy is displayed in Figure \ref{fig:ACCdys}. 

The results of this experiment indicate that the initial features selected by the $D_p$-distance criteria provide faster convergence to the maximum classification rate when compared to those selected by the BC criteria; however, as expected, as additional features are selected, both algorithms eventual converge to roughly the same level of performance. We purposefully restrict ourselves here to a very limited training set (300 samples per class) in order to evaluate the $D_p$-distance in a small $N$ setting. Next, we consider the same problem but with a variable number of training samples per class. The results of this experiment are presented in Table \ref{tab:results}. As the number of training instances increases, the classifier success rate increases for the $D_p$-based method, however it stays relatively flat for the BC-based method. For very small values of $N$, the bias/variance associated with the $D_p$-distance estimator seems to results in features that provide poorer separability when compared to the BC method. Given that the results of this estimator are asymptotic, this is expected. As the number of features increase, both the $D_p$ and BC algorithms converge to approximately the same value. 

\begin{table}
\centering
    \caption{Average classification accuracies (in percent) of top 10 features selected by $D_p$ and BC divergence}
\begin{tabular}{ c  c  c  c  c  c  c } 
\hline  
\hline
\bf{Number of} &\bf{Algorithm} & \multicolumn{5}{ c }{\bf{Number of Training Instances}} \\ 
\bf{Features} & \multicolumn{1}{c} {} & {\em 100} &  {\em 200} &  {\em 300} &  {\em 400} & {\em 500} \\ 
\hline 
 \multirow{2}{*}{10} & \pbox{25cm}{BC} & \textbf{86.88} & 86.93 & 87.61 & 87.98 & 87.22 \\
 & \pbox{25cm}{$D_p$} & 86.36 & \textbf{88.67} & \textbf{89.59} & \textbf{89.20} & \textbf{90.03}  \\
\multirow{2}{*}{15} & \pbox{25cm}{BC} & \textbf{90.84} & 90.46 & 90.51 & 91.69 & 90.88 \\
 & \pbox{25cm}{$D_p$} & 88.08 & \textbf{90.66} & \textbf{92.00} & \textbf{92.12} & \textbf{92.72}  \\ 
 \multirow{2}{*}{20} & \pbox{25cm}{BC} & \textbf{91.10} & \textbf{93.02} & \textbf{93.35} & \textbf{93.98} & 93.72 \\
 & \pbox{25cm}{$D_p$} & 89.28 & 92.15 & 93.20 & 93.41 & \textbf{94.21} \\
 \hline
\end{tabular}
\label{tab:results}
\end{table}

Next we would like to investigate the efficacy of the FS criteria (\ref{eq:crit2}) in a domain adaptation setting. We consider the same problem here - discriminating between healthy and dysarthric individuals; however now we train on data from one disorder and evaluate on data from another disorder. In order to partition the data into dissimilar training and test groups, we start by selecting 300 healthy instances for the training set and 300 (different) healthy instances for the test set. The rest of the training and test data is made up of 300 randomly selected samples from one of the four Dysarthria subtypes: Ataxic, ALS, Huntington's and Parkinson's. Each model is then evaluated on the test sets for each subtype not contained in the training set. 

Using each training set-test set combination, we generate feature subsets using the proposed selection algorithm, along with three competing algorithms that are used for comparison. The first algorithm we use for comparison is a standard forward selection algorithm based on the BC distance. This algorithm is used as a baseline for comparison, however because it assumes the training and test data come from the same distribution \cite{guorong1996}, we expect it to perform poorly relative to the other algorithms. Next we use the same Bhattacharyya FS algorithm, however we account for the separation in domains by using feature normalization, as described in \cite{Kinnunen2009}, prior to feature selection. We refer to this method as BC with feature normalization (BCFN). 

The final domain-invariant feature learning algorithm we compare against is based on Conditional Probability Models (CPM), as described in \cite{satpal2007domain}. This approach attempts to select a sparse mapping that maximizes an objective function that trades off between prediction algorithm performance and the distance between target and source distributions (controlled by a Lagrangian parameter $\lambda$). For classification, the logistic regression function is used and a penalization term is added to ensure that the mapping contains minimal contribution from features containing large differences between source and target data. For the specifics of the implementation, we refer the reader to \cite{satpal2007domain}. The same parameter settings are used here. Because this approach utilizes an optimization criteria involving a trade-off between the source-domain separation and the train-test separation, it resembles the proposed FS algorithm more closely than any other method proposed in the literature.

\begin{figure}
  \centering
  \includegraphics[width=0.45\textwidth]{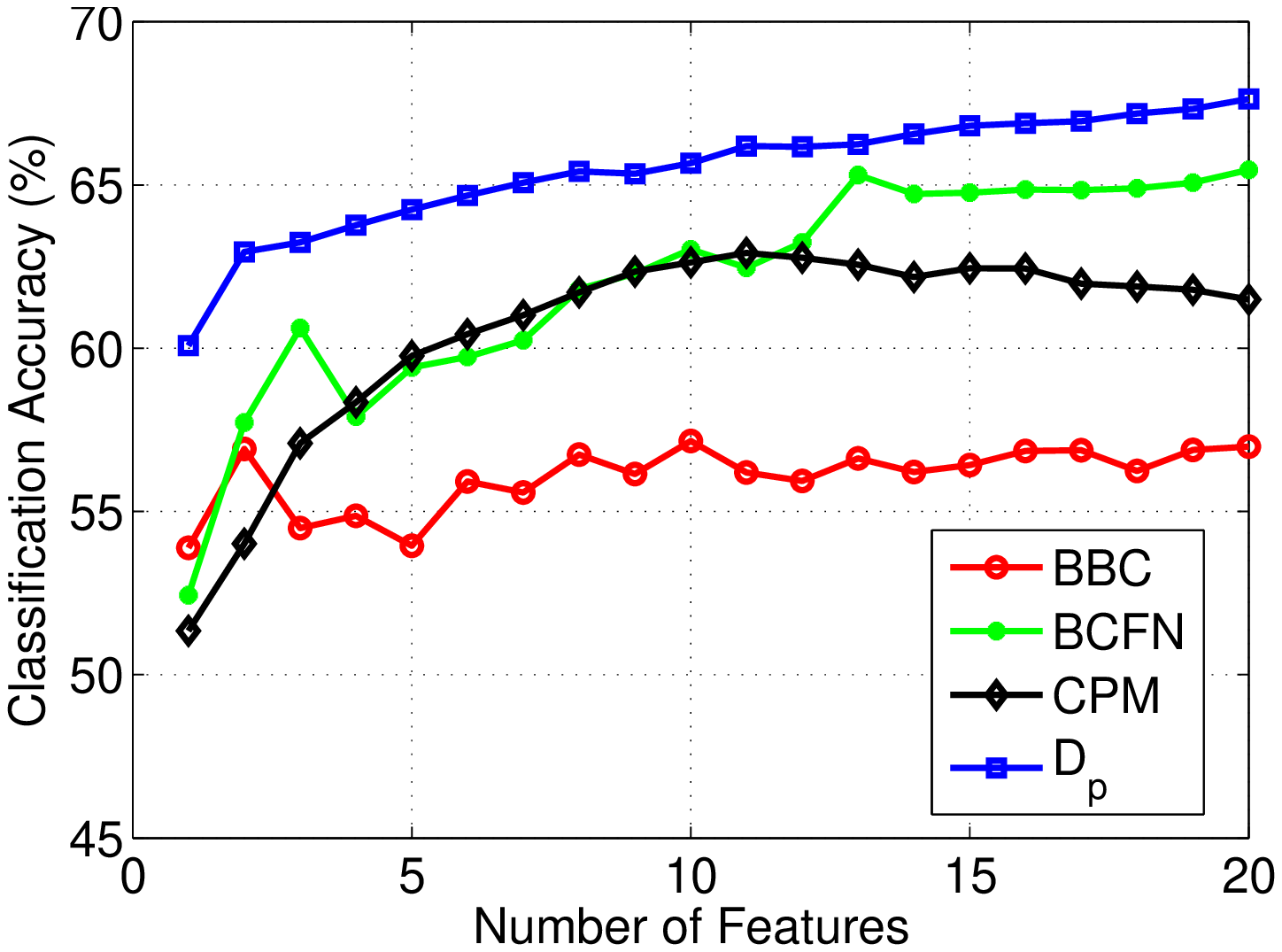}
  \caption{Average Classification Accuracy on foreign subtypes using reduced feature sets}
  \label{fig:ACCdysDA}
\end{figure}

We present the average classification accuracies yielded by the top 20 features from each FS algorithm for each train-test combination in Table \ref{tab:DysDAResults1}. The algorithm proposed in this paper acheived the highest classification accuracy in 8 of the 12 trials, while the BC algorithm scored the lowest 8 of 12 trials. The results clearly illustrate the importance of utilizing domain adaptation in this type of scenario; even an approach as simple as feature normalization yields roughly 8.5 \% higher classification accuracy on average. To observe the value of the lower-dimensional subsets generated by each algorithm, we average the accuracy across all twelve trials and display the accuracy as a function of the number of features in Figure \ref{fig:ACCdysDA}. We can see in this figure that the performance of the proposed algorithm consistently improves as additional features are added. Because the optimization criteria we have selected minimizes the upper bound on the error, the algorithm has a tendency to pick ``safe'' features; e.g. using this algorithm invariant features are preferred, even if they are less informative in the source domain.

\begin{figure}

\begin{center}
\subfloat[Source and target data using top domain-specific features]{
            \includegraphics[width=0.45\textwidth]{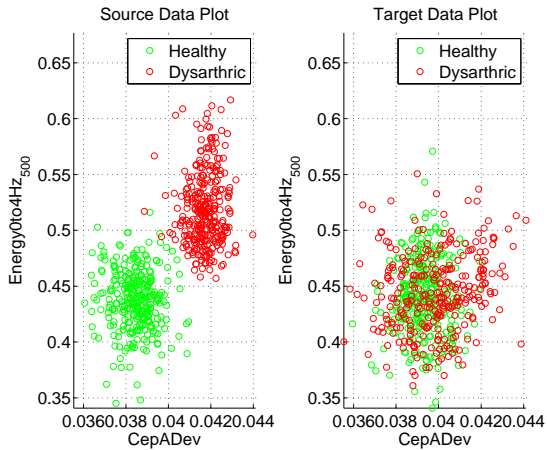}
            \label{fig:FRScat1}  
       }
       
\subfloat[Source and target data using top domain-invariant features]{
        \includegraphics[width=0.45\textwidth]{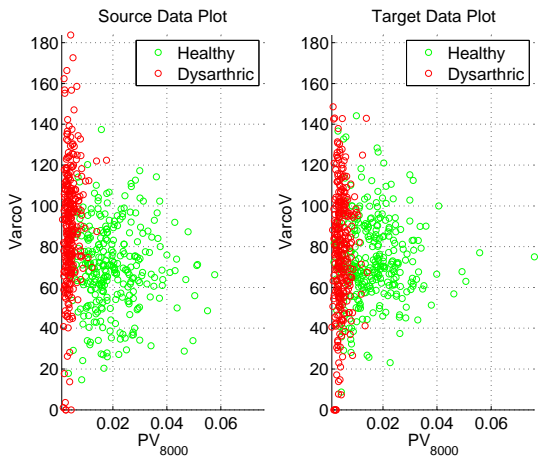}

            \label{fig:FRDAScat1}        
       }
       
       \caption{Low dimensional representation of datasets (Source
       Domain:ALS, Target Domain:Parkinson's).}
       \label{fig:Scatter Plots}
       \end{center}
\end{figure}

To better understand how DA helps us build robust models, we look at the top two features returned general and DA FS criterions proposed in this paper. Figure \ref{fig:FRScat1} displays the training and test data plotted across the top two features returned by the general FS criteria. We see that these two features represent a strong separation between the two classes in the training set, however this separation is not similarly represented in the test data, and as a result these features will not be beneficially for the target application. Figure \ref{fig:FRDAScat1} displays the data plotted against the top two features returned by the DA FS criteria. Even though the separation between classes in the training data isn't as noticable as in the features returned by the general criteria, both Dysarthria subtypes manifest themselves very similarly within this feature space, and as a result models built on them will generalize well between these two subtypes.

\begin{table}
\centering
    \caption{Classification accuracies of SVM classifier using the top 20 features returned by each feature selection method for each combination of training and test data}
\begin{tabular}{ c c  c c  c  c  c} 
\hline
\hline
Trial & Source & Target & BC & BCFN & CPM & $D_p$ \\ 
\hline 
1 & Ataxic & ALS & 56.50 & 73.28 & 75.82 & \textbf{76.22} \\ 
2 & Ataxic & Huntington's & 56.83 & 72.52 & 70.12 & \textbf{75.12} \\ 
3 & Ataxic & Parkinson's & 49.27 & 60.75 & 58.53 & \textbf{64.43} \\ 
4 & ALS & Ataxic & 52.95 & 66.35 & 54.68 & \textbf{67.15} \\ 
5 & ALS & Huntington's & 64.25 & \textbf{73.67} & 65.50 & 72.23 \\ 
6 & ALS & Parkinson's & 54.32 & 65.97 & 69.48 & \textbf{73.60} \\ 
7 & Huntington's & Ataxic & 49.95 & \textbf{53.63} & 43.00 & 49.30 \\ 
8 & Huntington's & ALS & 63.40 & 64.12 & 63.17 & \textbf{73.00} \\ 
9 & Huntington's & Parkinson's & 59.48 & 62.22 & 69.73 & \textbf{76.03} \\ 
10 & Parkinson's & Ataxic & 41.13 & \textbf{55.65} & 42.15 & 48.23 \\ 
11 & Parkinson's & ALS & 62.10 & 66.30 & 61.25 & \textbf{67.35} \\ 
12 & Parkinson's & Huntington's & \textbf{73.67} & 71.12 & 64.47 & 68.98 \\ 
\hline
\end{tabular}
\label{tab:DysDAResults1}

\end{table}          

\section{Conclusion}
In this paper we showed that a nonparametric $f$-divergence bounds the Bayes classification error rate for two scenarios: the case where  training and test data come from the same distribution and the case where training and test data come from different distributions. For the first case, we show that the bound is tighter than the commonly used Bhattacharyya bound on the Bayes error. Our experimental results confirm the theoretical findings - when used as a feature selection criterion in a pathological speech classification problem, the $D_p$-distance yields an improved classification rate with fewer features as compared against popular alternatives. 

Future work revolves around analyzing the estimator of the $D_p$-distance. In particular, understanding the convergence properties of the estimator as a function of the sample size and data dimension will yield insight into the fidelity of the estimation for any given data set. Furthermore, characterizing the bias and variance of this estimator may allow us to apply ensemble estimator methods of \cite{moon2014multivariate} to improve estimation accuracy for high dimensional feature space.

\appendices
 \section{Proof of Theorem \ref{eqn:div_HP}}
By combining Eq. (\ref{eqn:div_HP}) and (\ref{eqn:defineA}) from the text we can rewrite 
 \begin{align} 
 D_p & =  \frac{1}{4pq} [1 - 4pq A_p(f,g) - (p-q)^2] \\
  & =  \frac{1- (p-q)^2}{4pq}  - A_p(f,g) \\
   & = 1 - A_p(f,g), 
   \label{eqn:thm1proofDA}
 \end{align}
 where
 \begin{equation} \label{eqn:th1proof1}
 A_p = \int \frac{f(\mathbf{x})g(\mathbf{x})}{pf(\mathbf{x})+qg(\mathbf{x})} d\mathbf{x}.
 \end{equation}
 
 From Theorem 2 in \cite{henze1999multivariate}, we know that as $N_f \rightarrow \infty$ and $N_g \rightarrow \infty$ in a linked manner such that $\frac{N_f}{N_f+N_g}\rightarrow p$ and $\frac{N_g}{N_f+N_g}\rightarrow q$,
 \begin{equation} \label{eqn:thm1proofeq2}
 \frac{\mathcal{C}(f,g)}{N_f + N_g} \rightarrow 2pq A_p(f,g),
 \end{equation}
 almost surely.
 
 Combining the asymptotic relationship in Eq. (\ref{eqn:thm1proofeq2}) with the results from Eq. (\ref{eqn:thm1proofDA}), we see that
 \begin{equation}
  1- \mathcal{C}(f,g) \frac{N_f + N_g}{2N_fN_g} \rightarrow D_p(f,g),
 \end{equation} 
 almost surely as $N_f \rightarrow \infty$ and $N_g \rightarrow \infty$ in a linked manner such that $\frac{N_f}{N_f+N_g}\rightarrow p$ and $\frac{N_g}{N_f+N_g}\rightarrow q$.
 
 \section{Proof of Theorem \ref{thm:BER}}
  We begin with the realization that the Bayes error rate can be expressed in terms of the total variation (TV) distance between distributions \cite{kailath1967divergence}:
  \begin{equation}
  \epsilon^{Bayes}=\frac{1}{2}-\frac{1}{2}\int |pf(\mathbf{x})-qg(\mathbf{x})|d\mathbf{x}.
  \label{eq:BER1}
  \end{equation}  
  Next, we show that we can bound the TV distance from above and below using $\tilde{D}_p$:
   \begin{subequations}
   \begin{align}
   \tilde{D}_p & = 1-4pqA_p(f,g) \\
    &= 1 - 4pq\int \frac{f(\mathbf{x})g(\mathbf{x})}{pf(\mathbf{x})+qg(\mathbf{x})} d\mathbf{x}\\
  &\begin{aligned}= &\int \left[pf(\mathbf{x})+qg(\mathbf{x}) \right]d\mathbf{x} \\&- 4pq\int \frac{f(\mathbf{x})g(\mathbf{x})}{pf(\mathbf{x})+qg(\mathbf{x})} d\mathbf{x}\end{aligned}\\
                   &= \int \frac{[pf(\mathbf{x})+qg(\mathbf{x})]^2-4p q f(\mathbf{x})g(\mathbf{x})}{pf(\mathbf{x})+qg(\mathbf{x})} d\mathbf{x} \\
                   &= \int \frac{pf(\mathbf{x})^2+qg(\mathbf{x})^2-2p q f(\mathbf{x})g(\mathbf{x})}{pf(\mathbf{x})+qg(\mathbf{x})} d\mathbf{x} \\
                   &= \int \frac{[pf(\mathbf{x})-qg(\mathbf{x})]^2}{pf(\mathbf{x})-qg(\mathbf{x})} d\mathbf{x} \\
                   &= \int \left|pf(\mathbf{x})-qg(\mathbf{x})\right| \frac{\left|pf(\mathbf{x})-qg(\mathbf{x})\right|}{pf(\mathbf{x})+qg(\mathbf{x})}  d\mathbf{x} \label{eq:dist1f}.
  \end{align}
  \label{eq:dist1}
   \end{subequations}
   
   Since
     
   \begin{equation}
   \frac{\left|pf(\mathbf{x})-qg(\mathbf{x})\right|}{pf(\mathbf{x})+qg(\mathbf{x})}   \leq 1 \; \; \; \; \mathrm{for} \; \mathrm{all } \; \mathbf{x},
   \end{equation}
   we can simplify (\ref{eq:dist1f}) to 
   
   \begin{equation}
   1-4pqA_p(f,g) \leq \int \left|pf(\mathbf{x})-qg(\mathbf{x})\right|  d\mathbf{x}.
   \label{eq:A1}
   \end{equation}
   This provides a lower bound on the TV distance based on $\tilde{D}_p$. In order to derive the upper bound we begin with
  
   \begin{subequations}
   \begin{align}
   D_{\mathrm{TV}}(f,g)&=\int \left|pf(\mathbf{x})-qg(\mathbf{x})\right|  d\mathbf{x} \\
     &= \int \left|pf(\mathbf{x})-qg(\mathbf{x})\right| \frac{\sqrt{pf(\mathbf{x})+qg(\mathbf{x})}}{\sqrt{pf(\mathbf{x})+qg(\mathbf{x})}} d\mathbf{x}\\
   &\begin{aligned}&\leq \sqrt{\int \left(\frac{pf(\mathbf{x})-qg(\mathbf{x})}{\sqrt{pf(\mathbf{x})+qg(\mathbf{x})}}\right)^2 d\mathbf{x}} \\ &\times \cancelto{1}{\sqrt{\int \left(\sqrt{pf(\mathbf{x})+qg(\mathbf{x})}\right)^2 d\mathbf{x}}} \end{aligned}  \\
   &\leq  \sqrt{\tilde{D}_{p}(f,g)} \label{eq:TVbound2}. 
    \end{align}
    \end{subequations}
  
  By combining the inequalities in (\ref{eq:A1}) and (\ref{eq:TVbound2}) with the relationship in  (\ref{eq:BER1}), we see that we can bound the BER by
  
  \begin{equation}
  \frac{1}{2}-\frac{1}{2}\sqrt{\tilde{D}_{p}(f,g)} \leq \epsilon^{Bayes} \leq \frac{1}{2}-\frac{1}{2}\tilde{D}_{p}(f,g). 
  \end{equation}

 \section{Proof of Theorem \ref{thm:HPChernoff}}     
 
By the geometric vs harmonic mean inequality, 
\begin{equation}
f(\mathbf{x})^q g(\mathbf{x})^p \geq \frac{f(\mathbf{x})g(\mathbf{x})}{pf(\mathbf{x})+qg(\mathbf{x})}. 
\end{equation}
It immediately follows that  $A_p(f,g) \leq \int f(\mathbf{x})^q g(\mathbf{x})^p$, a scaled Chernoff information function. Thus,
 \begin{equation}
A_p(f,g) \leq \int f(\mathbf{x})^q g(\mathbf{x})^p.
\label{eq:qpbound}
\end{equation}

 \section{Proof of Theorem \ref{thm:tightness}} 
 
 For equiprobable classes ($p=q=\frac{1}{2}$) The upper and lower bounds on the Bayes error rate based on the Bhattacharyya distance are defined by \cite{kailath1967divergence}
 \begin{equation}
  \frac{1-\sqrt{1-BC^2(f,g)}}{2} \leq \epsilon^{Bayes} \leq \frac{BC(f,g)}{2},
  \label{eq:BB}
  \end{equation}
  where 
  \begin{equation}
  BC(f,g)=\int \sqrt{f(\mathbf{x})g(\mathbf{x})}d\mathbf{x}.
  \end{equation}
  
 To show that the $\tilde{D}_{\frac{1}{2}}$ bound upper bound is tighter than the Bhatacharyya bound we must show that $A_{\frac{1}{2}}(f,g) \leq BC(f,g)$. It is clear that this is the case from Theorem \ref{thm:HPChernoff}. For the $\tilde{D}_{\frac{1}{2}}$ lower bound to be tighter,  $BC^2(f,g)$ must be less than equal to $A_{\frac{1}{2}}(f,g)$. We show this to be true using the Cauchy-Schwartz inequality:
 
 \begin{subequations}
 \begin{align}
         BC^2(f,g)&=\left[\int \sqrt{f(\mathbf{x})g(\mathbf{x})}\right]^2 \\
          &=\left[ \int \frac{\sqrt{f(\mathbf{x})g(\mathbf{x})}}{\sqrt{\frac{1}{2}(f(\mathbf{x})+g(\mathbf{x}))}}\sqrt{\frac{1}{2}(f(\mathbf{x})+g(\mathbf{x}))}d\mathbf{x}\right]^2 \\
         &\leq \int \frac{f(\mathbf{x})g(\mathbf{x})}{\frac{1}{2}(f(\mathbf{x})+g(\mathbf{x}))}d\mathbf{x} \cancelto{1}{\int\frac{1}{2}(f(\mathbf{x})+g(\mathbf{x}))d\mathbf{x}} \\
          & = A_{\frac{1}{2}}(f,g).
 \end{align}
 \end{subequations}
 Combining both bounds, we see that
 \begin{equation*}
 \begin{aligned}
  \frac{1}{2}-\frac{1}{2}\sqrt{1-BC^2(f,g)} \leq & \frac{1}{2} - \frac{1}{2}\sqrt{\tilde{D}_{\frac{1}{2}}(f,g)}  \\   
 \leq \epsilon^{Bayes} \leq   \frac{1}{2} - & \frac{1}{2}\tilde{D}_{\frac{1}{2}}(f,g) \leq \frac{1}{2}BC(f,g).
  \end{aligned}
  \label{eq:thm3}
  \end{equation*}
 
 \section{Proof of Theorem \ref{thm:DAbound}} 
 
 The proof begins in the same fashion as the result in \cite{ben2010theory} and then diverges. 
 \begin{subequations}
 \begin{align}
 \epsilon_{\mathrm{T}} (h, y_{\mathrm{T}}) = &   \epsilon_{\mathrm{T}}  (h, y_{\mathrm{T}}) + \epsilon_{\mathrm{S}}  (h, y_{\mathrm{S}})  - \epsilon_{\mathrm{S}}  (h, y_{\mathrm{S}})  \\ 
 & + \epsilon_{\mathrm{S}} (h, y_{\mathrm{T}}) - \epsilon_{\mathrm{S}} (h, y_{\mathrm{T}}) \notag \\
   \leq &  \epsilon_{\mathrm{S}}(h, y_{\mathrm{S}})   +  |\epsilon_{\mathrm{S}} (h, y_{\mathrm{T}}) - \epsilon_{\mathrm{S}} (h, y_{\mathrm{S}})| \\
   & +  |\epsilon_{\mathrm{T}}(h, y_{\mathrm{T}}) - \epsilon_{\mathrm{S}} (h, y_{\mathrm{T}})| \notag \\
  \leq   &  \epsilon_{\mathrm{S}} (h, y_{\mathrm{S}})  + \mathbf{E}_{f_{\mathrm{S}}(\mathbf{x})}   [|y_S - y_T |] 
  \\ & + \Bigl \lvert \int f_{\mathrm{T}}(\mathbf{x}) |h(\mathbf{x}) - y_T| d\mathbf{x} \notag \\
  & \ \ \ \ \ - \int f_{\mathrm{S}}(\mathbf{x}) |h(\mathbf{x}) - y_T| d\mathbf{x} \Bigl \lvert \notag \\
  \leq &   \epsilon_{\mathrm{S}} (h, y_{\mathrm{S}}) + \mathbf{E}_{f_{\mathrm{S}}(\mathbf{x})}   [|y_S - y_T |]  \\
  & + \int | f_{\mathrm{T}}(\mathbf{x}) -f_{\mathrm{S}}(\mathbf{x}) | |h(\mathbf{x}) - y_T| d\mathbf{x}   \notag \\
  \leq   & \epsilon_{\mathrm{S}}(h, y_{\mathrm{S}})   + \mathbf{E}_{f_{\mathrm{S}}(\mathbf{x})}   [|y_S - y_T |] \label{finalbound} \\
  &+ \int | f_{\mathrm{T}}(\mathbf{x}) -f_{\mathrm{S}}(\mathbf{x}) |  d\mathbf{x} \notag    
 \end{align}
 \end{subequations}
 
 In (\ref{finalbound}), we identify an upper bound on the target error expressed using the TV distance between source and target distributions. Using (\ref{eq:TVbound2}) this can be expressed in terms of $\tilde{D}_{\frac{1}{2}}$:
 \begin{equation}
 \begin{aligned}
 \epsilon_T(h,y_T) & \leq \epsilon_S(h,y_S)  +E\{|y_S-y_T|\}   \\  
 &+2\sqrt{\tilde{D}_{\frac{1}{2}}(f_T,f_S)} \label{eq:DAbound}
 \end{aligned}
 \end{equation}

 %

\bibliographystyle{IEEEtran}
\bibliography{IEEEabrv,References}

\end{document}